\documentclass[usenatbib]{mn2e}

\usepackage{color}
\usepackage{graphicx}
\usepackage{bbold}
\usepackage{subfigure}
\usepackage{breqn}
\usepackage{threeparttable}
\usepackage{wasysym}
\usepackage{hyperref}
\hypersetup{pdfauthor=Judit Szul\'agyi}
\hypersetup{backref=true, pagebackref=true, hyperindex=true, breaklinks=true,colorlinks=true,urlcolor=blue, linkcolor=blue,  citecolor=blue,pagecolor=red, bookmarks=true, bookmarksopen=true}

\newcommand{\jup}{{\textrm{\small \jupiter}}}

\title[Shocking Hot Surfaces on Circumplanetary Disks]{Thermodynamics of Giant Planet Formation: Shocking Hot Surfaces on Circumplanetary Disks}
\author[Szul\'agyi \& Mordasini]{J. Szul\'agyi$^{1}$\thanks{E-mail:
judit.szulagyi@phys.ethz.ch} \& C. Mordasini$^{2}$\\
$^{1}$ETH Z\"urich, Institute for Astronomy, Wolfgang-Pauli-Strasse 27, CH-8093, Z\"urich, Switzerland\\
$^{2}$Physikalisches Institut, University of Bern, Sidlerstrasse 5, CH-3012 Bern, Switzerland\\}
\begin{document}

\date{Accepted XX. Received XX; in original form XX}

\pagerange{\pageref{firstpage}--\pageref{lastpage}} \pubyear{2016}

\maketitle

\label{firstpage}

\begin{abstract}
The luminosity of young giant planets can inform about their formation and accretion history. The directly imaged planets detected so far are consistent with the ``hot-start" scenario of high entropy and luminosity. If nebular gas passes through a shock front before being accreted into a protoplanet, the entropy can be substantially altered. To investigate this, we present high resolution, 3D radiative hydrodynamic simulations of accreting giant planets. The accreted gas is found to fall with supersonic speed in the gap from the circumstellar disk's upper layers onto the surface of the circumplanetary disk and polar region of the protoplanet. There it shocks, creating an extended hot supercritical shock surface. This shock front is optically thick, therefore, it can conceal the planet's intrinsic luminosity beneath. The gas in the vertical influx has high entropy which when passing through the shock front decreases significantly while the gas becomes part of the disk and protoplanet. This shows that circumplanetary disks play a key role in regulating a planet's thermodynamic state. Our simulations furthermore indicate that around the shock surface extended regions of atomic -- sometimes ionized -- hydrogen develop. Therefore circumplanetary disk shock surfaces could influence significantly the observational appearance of forming gas-giants.
\end{abstract}


\section{Introduction}
Giant planets are thought to form either via core-accretion \citep{Pollack96} or gravitational instability scenario \citep{Boss97}. To get a handle on which formation scenario led to an observed gas giant, the post-formation entropy of the planet was initially thought to distinguish between the two cases \citep{Burrows97,Marley07}. Traditionally,  planets formed by core accretion were thought to have a low luminosity and entropy ($\la$9.5 $\mathrm{k_B}$/baryon) -- corresponding to the so called ``cold-start" scenario -- whereas gravitational instability was thought to lead to giant planets having a high luminosity and entropy -- the ``hot-start" scenario ($\ga$9.5 $\mathrm{k_B}$/baryon, \citealt{Marley07}). Recent studies, however, pointed out that the situation is more complex. The entropy of the planets is affected by whether the accretion of gas onto the planet happens through a supercritical shock front, as indicated by one-dimensional spherically symmetric models \citep{Marley07}. If the gas forming the planet passes through such a entropy-reducing shock where a significant part of the accretion luminosity can be radiated, the planet itself will have a low entropy, consistent with the ``cold-start" scenario, regardless which formation mechanism builds the gas giant \citep{Mordasini12}. Moreover, the mass of the solid planetary core can alter the post-formation entropy of the planet as well, with higher mass cores leading to hotter planets \citep{Mordasini13,bodenheimerdangelo2013}. Finally, \citet{OM16} showed that the presence of a circumplanetary disk around the planet will funnel hot gas to the planet that can inflate the outer layers of the gas giant, enhancing the planet's entropy. 

Observations of directly imaged planets allow to measure the luminosities of young \citep[e.g.,][]{Marois08,lagrangebonnefoy2010} and recently also of still forming embedded planets \citep[e.g.,][]{KI12,Quanz15}. This makes it possible to estimate their entropies \citep{MC14} and conclude whether they are consistent with the ``cold-start'' or ``hot-start" scenarios. A handful of directly imaged gas giants is available for study today, and most seem to be luminous and consistent \citep{MC14,OM16} with the ``hot-start" scenario (which could in principle also be an observational bias as fainter planets are more difficult to detect). However,  gravitational instability as a formation mechanism appears unlikely in several of these cases due to various factors, such as the rather small semi-major axis or a rather low circumstellar disk mass \citep{FR13}. Furthermore, the luminosity estimations of forming embedded planets from direct imaging observations can be contaminated by the luminosity of the circumplanetary disk or accretion \citep{Zhu15,Szulagyi16}, enhancing the observed overall luminosity, that can make the planet look like a ``hot-start" case. The observation that the currently known directly observed planets seem to be consistent with a ``hot-start" scenario, but some of them may formed via core accretion (e.g., $\beta$ Pic b, \citealt{mordasinimolliere2015}) indicates that solely the low versus high entropy state of an observed gas giant cannot conclusively distinguish between the two planet formation scenario. This is partially due to the lack of theoretical studies of the thermodynamics of giant planet formation which predict the post-formation entropy based on multi-dimensional, radiation-hydrodynamical simulations. 

In this paper we therefore present a thermodynamical study based on 3D, radiative hydrodynamic simulations of forming gas giants with various masses embedded in circumstellar disks. The planets form circumplanetary disks or circumplanetary envelopes around them depending on the gas temperature in the planet vicinity \citep{Szulagyi16}. As described e.g. in \citet{Szulagyi14}, the accretion of the gas happens from the vertical direction through the planetary gap. This is because the top layers of the circumstellar disk try to close the gap opened by the giant planet, and, as gas enters the gap, it falls nearly freely onto the circumplanetary disk's surface and onto the polar regions of the protoplanet. The gas shocks at this surface, and then becomes the part of the disk, where it eventually spirals down to the planet. In this work we study the change of important thermodynamic quantities like the entropy or ionization state of the gas as it passes through the shock front.

\section{Methods}
\label{sec:methods}

Our study is based on three-dimensional, radiative, grid-based hydrodynamic simulations with the JUPITER code \citep{Borro06, Szulagyi14,Szulagyi16}, developed by F. Masset \& J. Szul\'agyi. This code is based on a shock-capturing Godunov's method using Riemann-solvers and has nested meshes, which allow to zoom onto the planet's vicinity with high resolution. The radiative module is based on the flux limited diffusion approximation \citep[e.g.][]{Commercon11}, and calculates temperatures where viscous heating and radiative cooling is included. The opacity of the gas and dust is taken into account through the \citet{BL94} opacity table. This means that despite of the purely gas hydrodynamic simulations, the impact of the dust on the temperature is included through the dust opacities and a fixed dust-to-gas ratio of 1\%. More details about the radiative code and the simulations is given in \citet{Szulagyi16} and in a future paper for the 3-10 $\mathrm{M_{\jup}}$ gas giants. Here we only mention the most important characteristics.  

The coordinate system is spherical, centered onto the star, and covers the entire circumstellar disk radially between 2.0 and 12.4 AU  with an initial surface density $\Sigma=\Sigma_0(\frac{r}{a})^{-0.5}$ with $\Sigma_0=2.22 \times 10^{3} \rm{kg/m^2}$ ($a$ being the semi-major axis, 5.2 AU). This density was chosen to be close to the Minimum Mass Solar Nebula (\citealt{Hayashi}). Only the lower half of the circumstellar (and circumplanetary) disks are simulated, assuming symmetry to the disk mid-plane. The equation of state (hereafter, EOS) in the code is the ideal gas EOS with an adiabatic index $\gamma$ equal to 1.43, such that $P=(\gamma-1)\epsilon$, where $\epsilon$ is the internal energy of the gas ($\epsilon=\rho C_v T$), and $P$ is the pressure. The mean molecular weight is 2.3, corresponding to the solar value. Due to the fixed adiabatic exponent and mean molecular weight, the ionization and dissociation of hydrogen and helium is not taken into account. The lack of these mechanisms could alter the predicted temperatures, entropies, and compositions, therefore our computations are only estimates found with an idealized EOS. Due to the extensive computation time needed for the high-resolution radiative simulations presented here, and due to the limitations of the solver, we had to use in this work an ideal EOS.

\begin{figure*}
\includegraphics[width=18cm]{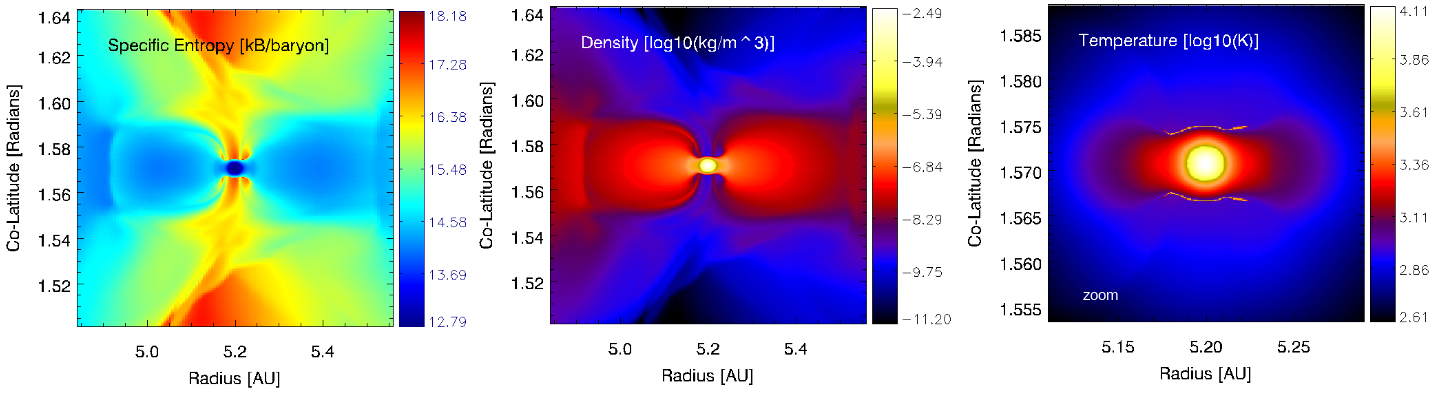}
\caption{Vertical slices of the specific entropy $S$ (Eq. \ref{eq::Sackur}), volume density $\rho$ and zoomed-in temperature $T$ for the planet of 3 $\mathrm{M_{\jup}}$. For $S$ and $\rho$ the entire circumplanetary disk is shown. In the left-hand panel, one sees that the gas falling from the vertical direction has a high entropy, then shocks on the surface of the circumplanetary disk, leading to a strong entropy reduction for the accreted gas. The circumplanetary disk has a significantly lower entropy, and the minimal entropy can be found in a small, spherical dark-blue envelope around the planet's location. In the right-hand panel, part of the hot shock surface on the circumplanetary is visible as a razor-thin layer (the Zeldovich spike). We show only the 3 $M_{\jup}$ planet, but the 5 and 10 $M_{\jup}$ planets look qualitatively the same.}
\label{fig::entr}
\end{figure*}

We performed simulations with 1, 3, 5, and 10 $\mathrm{M_{\jup}}$ planets for $\sim220$ orbits until steady state has been reached. A constant kinematic viscosity was applied with value of $10^{-5} a^2\Omega_p$, where $a$ is the semi-major axis, $\Omega_p$is the orbital frequency. This viscosity corresponds to a value of $\alpha=0.004$ at 5.2 AU. Due to the nested meshes centered on the planet, the resolution changes mesh by mesh, with each level of refinement doubling the resolution. This way the highest level of resolution (reached on level 6) was $7.5\times10^{-4}\,$AU$\,=\,0.8\,\mathrm{d_{\jup}}$ where $\mathrm{d_{\jup}}$ is the diameter of Jupiter. In the simulations the planet is treated as a point-mass in the corner of 8 cells. In other words there is only a gravitational potential well, no sphere is modeled for the giant planet. The smoothing lengths were \ $4.4 \times 10^{-3}$, $8.8\times 10^{-3}$, $8.8\times 10^{-3}$ AU, and  $1.8\times 10^{-2}$ AU for the 1, 3, 5, and 10 $\mathrm{M_{\jup}}$ planets, respectively. As it was shown in \citet{Szulagyi16}, in the 1 $\mathrm{M_{\jup}}$  simulation the gas is too hot in the planet's vicinity to collapse into a circumplanetary disk. Instead, a pressure supported spherical envelope formed around the planet. In this case, there is no shock front on the surface of the envelope where the vertical influx hits it. We show the entropies of this simulation for comparison, in order to distinguish between cases with and without supercritical shock fronts.

\section{Results}\label{sec:entropy}
To estimate the specific entropy ($S$) of the gas, we used four different approaches. All four give the specific entropy  based on the pressure and the temperature from the hydrodynamic simulations. The first expression we applied is the classical Sackur-Tetrode entropy formula for $\mathrm{H_2}$/He mixture with He mass fraction of 25\% \citep{MC14}:

\begin{equation}
S ($\rm{k_B}$/baryon)=9.6+\frac{45}{32}\ln (T/1600 \mathrm{K}) - \frac{7}{16} \ln(P/3\,\mathrm{bar})
\label{eq::Sackur}
\end{equation}

This expression does not include ionization/dissociation, such as the EOS used in the hydrodynamic code. The left-hand panel of Fig. \ref{fig::entr} shows the specific entropy -- calculated with the Sackur-Tetrode expression -- on a vertical slice through the 3 $M_{\jup}$ planet, zooming to the planet vicinity. The circumplanetary disk pops out with blue colors, which has clearly lower entropy than the supersonic vertical gas influx (red-yellow colors) which feeds the circumplanetary disk. The minimum entropy can be found in a spherical envelope around the planet, within the inner circumplanetary disk. In Fig. \ref{fig::entr} we also show the same vertical cut of the density and a further zoomed-in temperature map in the middle and the right-hand panels, respectively. On the temperature color-map one can see with bright yellow color part of the razor-sharp shock layer (the Zeldovich spike) right above the planet and on the surface of the inner circumplanetary disk. The shock however, is even more extended on the top layer of the circumplanetary disk, based on maps of the Mach number. The temperature just above and below Zeldovich spike is identical, meaning that this shock front is super-critical \citep{Vaytet2013} for all the different mass planets which are forming circumplanetary disks (3-10 $\mathrm{M_{\jup}}$ planets). In the case of the 1 Jupiter-mass simulation, as it was mentioned in the previous section, the circumplanetary gas is too hot to collapse into a disk, therefore an envelope is found (see Fig. 1 in \citealt{Szulagyi16}). Due to the extended envelope, the vertical influx does not hit it with large enough velocity (i.e. subsonically) to create a shock front on the surface of the envelope. 

\begin{figure}
\includegraphics[width=\columnwidth]{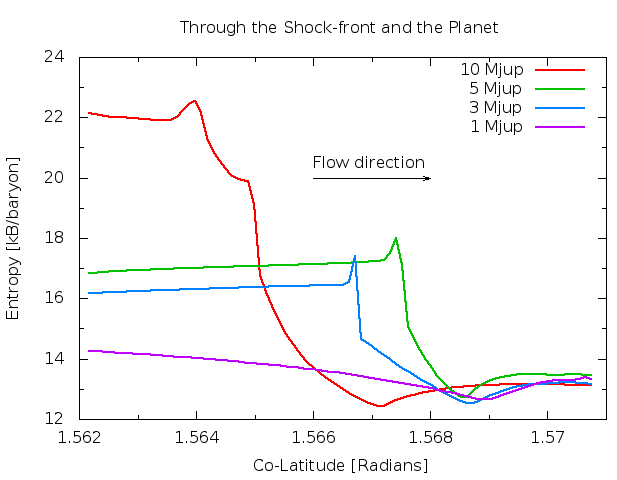}
\caption{1D vertical entropy profiles of the gas flow passing through the shock, right below (or above) the planet location. The rightmost value is at the midplane, the leftmost value is in the vertical influx. The peaks in the 3-10 $M_{\jup}$ simulations are the Zeldovich spikes at the supercritical shock front. The gas undergoes a strong entropy reduction ($> 3\mathrm{k_B}$/baryon) via radiative cooling while passing through the shock.}
\label{fig::shockprof_cdp}
\end{figure}

The second EOS we used was the Saumon-Chabrier-van Horn EOS developed for interiors of giant planets \citep{SCh95} using the same He/H mass fraction. The third way to calculate the specific entropies was using the CEA Gibbs minimizer \citep{mcbridegordon1993}. This approach, unlike the Sackur-Tetrode expression, calculates the specific entropy, mean molecular weight, and adiabatic index with the dissociation and ionization of hydrogen and helium taken into account. However, for comparison we also did a calculation with this code enforcing the molecularity and neutrality of the hydrogen-helium gas (only H$_{2}$ and He). This corresponds again to a situation close to the EOS used in the hydrodynamic simulations, except for the inclusion of the variation of the degrees of freedom with temperature.

The Fig. \ref{fig::shockprof_cdp} shows a 1D vertical profile of the entropy as the gas passes through the shock front right below the planet (vertical cut through radius\,=\,5.2 AU, azimuth\,=\,0.0). As mentioned in Sect. \ref{sec:methods}, only the lower half of the circumstellar disk is simulated, therefore the co-latitute ranges in this figure from 1.562 radians to 1.57079 ($\frac{\pi}{2}$), the latter being the midplane value. The entropies shown were calculated with CEA assuming only H$_{2}$ and He. One can observe that the gas lost significant amount of entropy ($>3\mathrm{k_B}$/baryon) when passing through the shock of the circumplanetary disk. The shock front itself is visible as a spike (the Zeldovich spike). It is obvious that in the vertical influx and at the Zeldovich spike the entropy is very high (17-23 $k_{\rm B}$/baryon) as we are dealing with low-density preheated gas falling from the top of the gap. The entropy is minimal at the planet location (around 13 $\mathrm{k_B}$/baryon for each planet). This shows that also in 3D, an accretion shock is found to play a crucial role in regulating the planet's thermodynamic state. Comparing the 3-10 $\mathrm{M_{\jup}}$ gas giants, it can be observed that with increasing mass, the stronger is the shock, and the higher is the radiative entropy reduction. This is visible from the decreasing value of the minimum of $S$ at some distance behind the shock. Also on the mid-plane (right-end of the figure) the 10 $\mathrm{M_{\jup}}$ simulation has the lowest entropy. In the case of the 1 $M_{\jup}$ planet, where there is only a circumplanetary envelope without any shock, the entropy almost steadily decreasing while approaching the planet.

The shock $S_{\rm shock}$ and post-shock $S$ entropies (measured in and immediately after the Zeldovich spike, respectively) for the four different EOS can be found in Table \ref{tab::entropy}. The motivation to measure the entropy after the spike is to estimate the entropy of the gas which the planet will eventually accrete. Currently we find high, ``hot-start'' like entropies in the ideal gas EOS limit. The Saumon-Chabrier EOS also give very similar entropy values (within 0.1 $\mathrm{k_B}$/baryon) to the  CEA values, as expected. The Sackur-Tetrode expression gives somewhat higher (by 0.6-0.7 $\mathrm{k_B}$/baryon) specific entropies. In contrast to this, in the Zeldovich spike itself (and in a small envelope right around the planet's position), the entropy values differ significantly when ionization and dissociation is included in CEA relative to the neutral case, since here ionization and dissociation should happen. 

\begin{table*}
  \caption{Temperatures, post-shock, and shock entropy for different planet masses}
 \label{tab::entropy}
\begin{threeparttable}
  \begin{tabular}{p{1.4cm}p{1.3cm}p{1.4cm}p{0.9cm}p{1.3cm}p{1.7cm}p{2.2cm}p{2.0cm}p{1.7cm}}
  \hline
   Planet mass  $[\mathrm{M_{\jup}}]$ &  $\mathrm{T_{shock}}$ [K]  &  $\mathrm{P_{shock}}$ [$\rm{dyn/cm^2}$] &  $\mathrm{S}$ (ST)  &  $\mathrm{S}$ (SCvH) & $\mathrm{S}$ (CEA-ion)&   $\mathrm{S}$ (CEA-neutral) & $\mathrm{S_{shock}}$ (CEA-ion)& $\rm{S_{shock}}$(CEA-neutral)\\
 \hline
3.0   &   3893 & 0.42 &   15.39  &  14.73  & 14.66   &  14.66   & 28.39 &  17.40  \\
5.0   &   4429 & 0.19 &  15.74   &   15.14 & 15.05   &  15.05 &  29.33 & 18.00\\ 
10.0  &   8281 & 9.87e-5 & 17.51  &  16.85  & 16.77  & 16.77 &  62.69 &  22.53\\
 \hline
\end{tabular}
\begin{tablenotes}
\small{\item The entropies $S$ were measured immediately after the Zeldovich spike, and $S_{\rm shock}$ in the spike. ST denotes the Sackur-Tetrode expression, SCvH the Saumon-Chabrier-van Horn EOS, CEA-ion and CEA-neutral is the Chemical Equilibrium Code with ionization and dissociation and without. Entropies are in $\mathrm{k_B}$/baryon. $P$ is the pressure, $T$ stands for the temperature.}
\end{tablenotes}
\end{threeparttable}
\end{table*}

\begin{figure}
\includegraphics[width=\columnwidth]{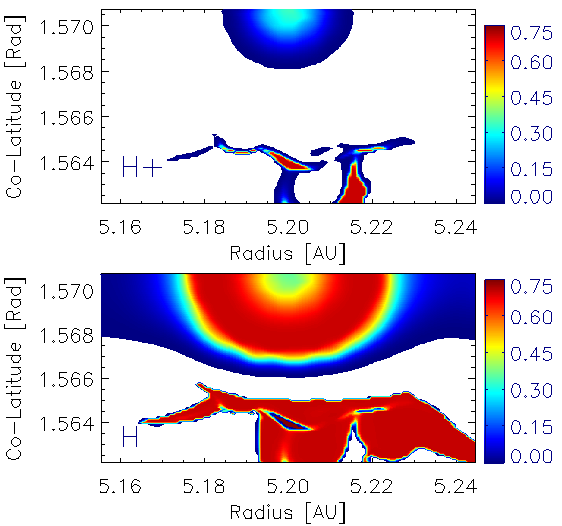}
\caption{The ionization (top) and dissociation (bottom) of hydrogen shown in mass fractions zoomed into the envelope around the planet, where these occur for the 10 $\mathrm{M_{\jup}}$ planet. Only the lower half of the disk is shown, so the mid-plane is at the top of the figures. In this case, ionization occurs at the shock front, which suggests that there is H-$\alpha$ emission. Dissociation only is found in the shock surfaces of 3 and 5 $\mathrm{M_{\jup}}$ gas giants.}
\label{fig::ion-diss}
\end{figure}

We also estimated with CEA the mass fractions of H (dissociation) and H+ (ionization) (Fig. \ref{fig::ion-diss}) for the 10 $\mathrm{M_{\jup}}$ planet. For all planetary masses the polar shock surface is so hot ($T>3800$ K), that the H$_2$ molecules are dissociated, and in the 10 Jupiter case ($T>8000$ K) hydrogen is even ionized. The presence of H+ at the shock surface assumes a detectable H-$\alpha$ emission from this extended (in comparison to a typical planetary radius of a few Jovian radii) region. This typical accretion tracer was recently detected around the planet, LkCa15b \citep{Sallum15}. We calculated the upper limit for H-$\alpha$ flux from the shock of the 10 $\mathrm{M_{\jup}}$ planet using Eq. 21 in \citet{Zhu15} and found $ 3\times10^{-2} \rm{L_{\odot}}$. For the planetary accretion, we applied the expression $\rm{L_{acc}}=GM_p\dot{M_p}/R_p$ and its scaling with $L_{\mathrm{H}\alpha}$ used in \citet{Sallum15}. The accretion rates to the planet calculated from our simulations are 9.8, 13.89, $5.7 \times 10^{-8} \mathrm{M_{\jup}/year}$ for the 3, 5, 10 $\mathrm{M_{\jup}}$ planets, which indicate $\sim$ 4 to 8 $\times10^{-5} \rm{L_{\odot}}$ for $\rm{L_{acc}}$. This translates to $\sim$ 4 to 7 $\times10^{-6} \mathrm{L_{\odot}}$ for H-$\alpha$ luminosity, an order of magnitude lower than what was measured in LkCa15b. Given that our H-$\alpha$ line flux upper limit is three orders of magnitude higher than our planet emitted line flux, it is also plausible that the shock of circumplanetary disk accounts for difference found with \citet{Sallum15}. Our results indicate, that by detecting the H-$\alpha$ emission, one cannot necessarily distinguish whether it tracks the accretion onto the circumplanetary disk or from the accreting planet itself.

When observing a forming embedded giant planet via direct imaging, one has to be careful from where the detected luminosity originates. As we see on the temperature color map of Fig. \ref{fig::entr}, part of the shock front on the surface of the circumplanetary disk and protoplanet is very luminous due to shock heating. This part of the shock front extends to $\sim$100-250 $\mathrm{R_{\jup}}$ in diameter for the 3 to 10 $\mathrm{M_{\jup}}$ planet simulations. This part is optically thick in our grey approximation, so it is possible that observationally this shock front is the surface the observations detect, at least in some wavelengths, rather than the actual protoplanet below. Because the luminosities of forming directly imaged planets are used to distinguish between ``hot-start'' and ``cold-start'' scenarios, and also to estimate the planetary mass, it is important to consider a possible contribution from the circumplanetary disk shock luminosity. This is in particular the case if it is not possible to distinguish observationally (spectroscopically) the origin of the radiation, e.g., because hard shock radiation first gets reprocessed in the surrounding disk and re-radiated at longer wavelength. This highlights that observationally, even intrinsically cold planets could look like ``hot-start'' planets during formation if they are surrounded by luminous shocks on the circumplanetary disk surface.  In conclusion, circumplanetary disk shock surfaces play a key role not only in regulating a planet's post-formation thermodynamic state, but also for the observational appearance of protoplanets during formation.

\section{Conclusions \& Discussions}

In this paper we present a study of the thermodynamics found in global three-dimensional radiative hydrodynamical simulations of embedded accreting giant planets of 1, 3, 5, 10 $\mathrm{M_{\jup}}$.  As described in detail in \citet{Szulagyi14}, the accretional gas flow from the circumstellar disk to the planet is the following. The gas acts to close the gap opened by the planet in the circumstellar disk, especially so in the high co-latitute regions. Gas enters the gap region, and then falls nearly freely in a vertical influx onto the circumplanetary disk and protoplanet. Because this vertical inflow is supersonic (MACH = 6.2, 8.1, 10.3 for the 3, 5, 10 $\mathrm{M_{\jup}}$ planets, respectively), it shocks on the surface of the circumplanetary disk and on the polar region of the protoplanet before becoming part of the disk and eventually reaching the protoplanet. In this work we showed that the gas undergoes a significant reduction of the specific entropy (typically more than 3 $\mathrm{k_B}$/baryon) while passing through the shock front that is found to be supercritical. The vertical influx has a very high entropy which after the shock in the disk mid-plane reaches a minimum value. We found that the circumplanetary disk consists of gas of significantly lower entropy than the vertical influx and the lowest entropies are found in a spherical small envelope around the planet within the inner parts of the circumplanetary disk. We conclude that shocks play a key role in regulating the post-formation entropy.

Because the shock front on the circumplanetary disk is hot, optically thick (in our grey-approximation), and this luminous region is extended (100-250 $\mathrm{R_{\jup}}$), it can contribute strongly to the bolometric luminosity of a directly imaged planet if the gas giant is still accreting. Therefore it is important to disentangle the luminosity of the shock front on the upper layer of the circumplanetary disk from the luminosity of the protoplanet itself beneath the shock surface. 
 
Our radiative hydrodynamic simulations compute temperatures taking into account radiative cooling with the inclusion of dust opacities. However, the use of ideal gas EOS does not take into account ionization and dissociation. This can lead to too high temperatures, and means that we cannot yet exactly predict the post-formation entropies. To estimate where dissociation and ionization could occur, we used the CEA code to determine potential H and H+ regions. We found that for 3-10 $\mathrm{M_{\jup}}$ planets dissociation  occurs in front of the polar shock surface indicating that non-ideal effect could be important. In the 10 Jupiter-mass simulation the shock surface is so hot to also produced H+, which means that there could be extended $H-\alpha$ emission from this region, which may be detectable, as seen in LkCa 15 b \citep{Sallum15}. Shock surfaces on the surface of the circumplanetary disk and the polar region of the protoplanet could therefore also be important for the observational appearance of forming giant planets.
\section*{Acknowledgments}

We are thankful to J. Owen, T. Guillot, G. Marleau for useful discussions and to the anonymous referee for improving this work. J. Sz. acknowledges the support from the ETH Post-doctoral Fellowship by the Swiss Federal Institute of Technology. This work has been in part carried out within the frame of the National Centre for Competence in Research ``PlanetS"  supported by  the  Swiss  National Science Foundation.  C.M. acknowledges the support from the Swiss National Science Foundation under grant BSSGI0$\_$155816 ``PlanetsInTime''. Computations have been done on the ``M\"onch" machine (Swiss National Comp. Cent.)

\label{lastpage}


\begin{thebibliography}{99}
\bibitem[\protect\citeauthoryear{Bell \& Lin}{1994}]{BL94} Bell, K.~R., \& Lin, D.~N.~C.\ 1994, ApJ, 427, 987 
\bibitem[\protect\citeauthoryear{Bodenheimer et al.}{2013}]{bodenheimerdangelo2013} Bodenheimer, P., D'Angelo, G., Lissauer, J.~J. et al.\ 2013, ApJ, 770, 120 
\bibitem[\protect\citeauthoryear{Boss}{1997}]{Boss97} Boss, A.~P.\ 1997, Science, 276, 1836 
\bibitem[\protect\citeauthoryear{Burrows et al.}{1997}]{Burrows97} Burrows, A., Marley, M., Hubbard, W.~B., et al.\ 1997, ApJ, 491, 856 
\bibitem[\protect\citeauthoryear{Commer{\c c}on et al.}{2011}]{Commercon11} Commer{\c c}on, B., Teyssier, R., Audit, E., et al.\ 2011, A\&A, 529, A35 
\bibitem[\protect\citeauthoryear{de Val-Borro et al.}{2006}]{Borro06} de Val-Borro, M., Edgar, R.~G., Artymowicz, P., et al.\ 2006, MNRAS, 370, 529
\bibitem[\protect\citeauthoryear{Forgan \& Rice}{2013}]{FR13} Forgan, D., \& Rice, K.\ 2013, MNRAS, 432, 3168 
\bibitem[\protect\citeauthoryear{Hayashi}{1981}]{Hayashi} Hayashi, C., 1981, Progress of Theoretical Physics Supplement, 70, 35 
\bibitem[\protect\citeauthoryear{Kraus \& Ireland}{2012}]{KI12} Kraus, A.~L., \& Ireland, M.~J.\ 2012, ApJ, 745, 5 
\bibitem[\protect\citeauthoryear{Lagrange et al.}{2010}]{lagrangebonnefoy2010} Lagrange, A.-M., Bonnefoy, M., Chauvin, G., et al.\ 2010, Science, 329, 57 
\bibitem[\protect\citeauthoryear{Marois et al.}{2008}]{Marois08} Marois, C., Macintosh, B., Barman, T., et al.\ 2008, Science, 322, 1348 
\bibitem[\protect\citeauthoryear{Marleau \& Cumming}{2014}]{MC14} Marleau, G.-D., \& Cumming, A.\ 2014, MNRAS, 437, 1378 
\bibitem[\protect\citeauthoryear{Marley et al.}{2007}]{Marley07} Marley, M.~S., Fortney, J.~J., Hubickyj, O., et al. \ 2007, ApJ, 655, 541 
\bibitem[\protect\citeauthoryear{McBride et al.}{1993}]{mcbridegordon1993} McBride, B., Gordon, S., \& Reno, M. 1993, NASA Technical Memorandum, 4513 
\bibitem[\protect\citeauthoryear{Mordasini et al.}{2012}]{Mordasini12} Mordasini, C., Alibert, Y., Klahr, H., \& Henning, T.\ 2012, A\&A, 547, A111 
\bibitem[\protect\citeauthoryear{Mordasini}{2013}]{Mordasini13} Mordasini, C.\ 2013, A\&A, 558, A113 
\bibitem[\protect\citeauthoryear{Mordasini et al.}{2015}]{mordasinimolliere2015} Mordasini, C., Molli{\`e}re, P., Dittkrist, K.-M., et al. \ 2015, Int. J. of Astrobiol., 14, 201 
\bibitem[\protect\citeauthoryear{Owen \& Menou}{2016}]{OM16} Owen, J.~E., \& Menou, K.\ 2016, ApJL, 819, L14 
\bibitem[\protect\citeauthoryear{Pollack et al.}{1996}]{Pollack96} Pollack, J.~B., Hubickyj, O., Bodenheimer, P., et al., 1996, Icarus, 124, 62 
\bibitem[\protect\citeauthoryear{Quanz et al.}{2015}]{Quanz15}  Quanz, S.~P., Amara, A., Meyer, M.~R., et al.\ 2015, ApJ, 807, 64 
\bibitem[\protect\citeauthoryear{Sallum et al.}{2015}]{Sallum15} Sallum, S., Follette, K.~B., Eisner, J.~A., et al.\ 2015, Nature, 527, 342 
\bibitem[\protect\citeauthoryear{Saumon et al.}{1995}]{SCh95} Saumon, D., Chabrier, G., \& van Horn, H.~M.\ 1995, ApJS, 99, 713 
\bibitem[\protect\citeauthoryear{Szul{\'a}gyi et al.}{2014}]{Szulagyi14} Szul{\'a}gyi, J., Morbidelli, A., Crida, A., \& Masset, F.\ 2014, ApJ, 782, 65
\bibitem[\protect\citeauthoryear{Szul{\'a}gyi et al.}{2016a}]{Szulagyi16} Szul{\'a}gyi, J., Masset, F., Lega, E., et al.\ 2016, MNRAS, 460, 2853 
\bibitem[\protect\citeauthoryear{Vaytet et al.}{2013}]{Vaytet2013} Vaytet, N., et al. \ 2013, J. of Quantitative Spectroscopy and Radiative Transfer, 125, 105 
\bibitem[\protect\citeauthoryear{Zhu}{2015}]{Zhu15} Zhu, Z.\ 2015, ApJ, 799, 16 
\end{thebibliography}
\end{document}